\documentclass[]{article}
\usepackage{enumerate}
\usepackage{indentfirst}
\usepackage{graphicx}
\usepackage{amsmath}
\title{\textbf{Fair Packet Scheduling in NoC}}
\author{\small Zhuang Wang, Xiao Lv, Mingyu Yan, Wei Yang, Ge Li \\ \small Institue of Computing Technology, Chinese Academy of Sciences}
\date{}
\begin{document}

\maketitle

\begin{abstract}
Interconnection networks of parallel systems are used for servicing traffic generated by different applications, often belonging to different users. When multiple traffic flows contend for channel bandwidth, the scheduling algorithm regulating the access to that channel plays a key role in ensuring that each flow obtains the required quality of service. Fairness is a highly desirable property for a scheduling algorithm. We show that using the Relative Fairness Bound as a fairness measure may lead to decrease in throughput and increase in latency. We propose an alternative metric to evaluate the fairness and avoid the drawback of Relative Fairness Bound.
\end{abstract}

\section{Introduction}
Interconnection networks of parallel systems are used for servicing traffic generated by different applications, often belonging to different users. When multiple traffic flows contend for channel bandwidth, the scheduling algorithm regulating the access to that channel plays a key role in ensuring that each flow obtains the required quality of service. Fairness is a highly desirable property for a scheduling algorithm. A variety of scheduling algorithms have been proposed to meet the demand of fairness. We will introduce some related works herein and discuss another metric different from current mean to measure the fairness.

\section{Fairness measure}
In order to define the fairness, a measure called Relative Fairness Bound (RFB) has been proposed in \cite{DDR}. M. Shreedhar et al. argue that it makes no sense to compare a flow that is not backlogged with one that is, because the former does not receive any service when it is not backlogged.

\newtheorem{definition}{Definition}
\begin{definition}
A flow is backlogged during an interval I of an execution if the queue for flow I is never empty during interval I.
\end{definition}
\begin{definition}
Let $FM(t_1,t_2)$ be the maximum, over all pairs of flows i, j that are backlogged in the interval $(t_1,t_2)$, of $(sent_i (t_1,t_2)/ f_i- sent_i(t_1,t_2)/ f_j)$. Define FM to be the maximum value of $FM(t_1,t_2)$ over all possible executions of the fair queuing scheme and all possible intervals $(t_1,t_2)$ in an execution.
\end{definition}
\begin{definition}
A service discipline is fair if FM is a small constant. In particular, $FM(t_1,t_2 )$ should not depend on the size of the interval $(t1, t2)$.
\end{definition}
This evaluation standard is also known as Relative Fairness Bound.

\section{Related work}
\subsection{Deficit Round Robin}
An algorithm for fairly servicing queues in a router call deficit round-robin (DRR) is designed in \cite{DDR}.

They assume that the quantities $f_i$, which indicate the share given to flow i, are specified as follows. They assume that each flow i is allocated $Q_i$  worth of bits in each round. Define $Q=Min_i(Q_i)$, the share $f_i$ allocated to flow i is simply $Q_i/Q$. Finally, since the algorithm works in rounds, they measure time in terms of rounds. A round is one round-robin iteration over the queues that are backlogged.

Packets coming in on different flows are stored in different queues. Let the number of bytes sent out for queue i in round k be $bytes_i(k)$. Each queue i is allowed to send out packets in the first round subject to the restriction that $bytes_i(1)<Q_i$. If there are no more packets in queue i after the queue has been serviced, a state variable called $DC_i$ is reset to zero. Otherwise, the remaining amount $(Q_i-bytes_i(k))$ is stored in the state variable $DC_i$. In subsequent rounds, the amount of bandwidth usable by this flow is the sum of $DC_i$ of the previous round added to $Q_i$.

To avoid examining empty queues, they keep an auxiliary list ActiveList that is a list of indices of queues that contain at least one packet. Whenever a packet arrives to a previously empty queue i, i is added to the end of ActiveList. Whenever index i is at the head of ActiveList, the algorithm services up to $(Q_i+DC_i)$ worth of bytes from queue i ; if at the end of this service opportunity, queue i still has packets to send, the index i is moved to the end of ActiveList ; otherwise, $DC_i$ is set to zero and index i is removed from ActiveList. Pseudo-code for this algorithm is shown in Fig. 1.
\begin{figure}[ht]
\centering
\includegraphics[width=8cm]{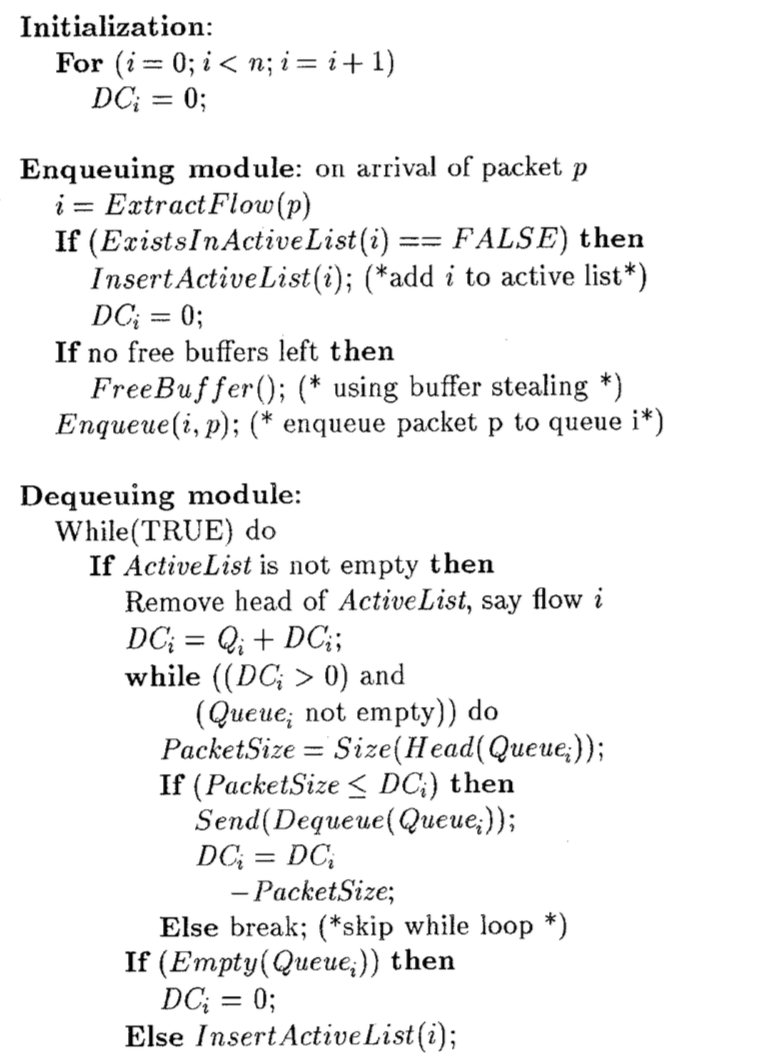}
\caption{Pseudo-code for DRR}
\end{figure}
Enqueue (), Dequeue () are standard Queue operators. They use a list of active flows, ActiveList, with standard operations like lnsertActiveList, which adds a now index to the end of the active list. FreeBuffer () frees a butter from the flow with the longest queue using McKinney’s buffer stealing. $Q_i$ is the quantum allocated to $Queue_i$. $DC_i$ contains the bytes that $Queue_i$ did not use in the previous round.

\subsection{Elastic Round Robin}
DRR is a simple, fair, efficient and easily implementable scheduling discipline, while the information of packet size is needed. However, when a scheduler uses the size of a packet to make its decision, it cannot be readily adapted to the unique requirements of wormhole switching. Elastic Round Robin \cite{ERR} satisfy the need of fairness without the knowledge of packet size, therefore it can be used in both Internet routers and wormhole switches.
\begin{figure}[!ht]
\centering
\includegraphics[width=8cm]{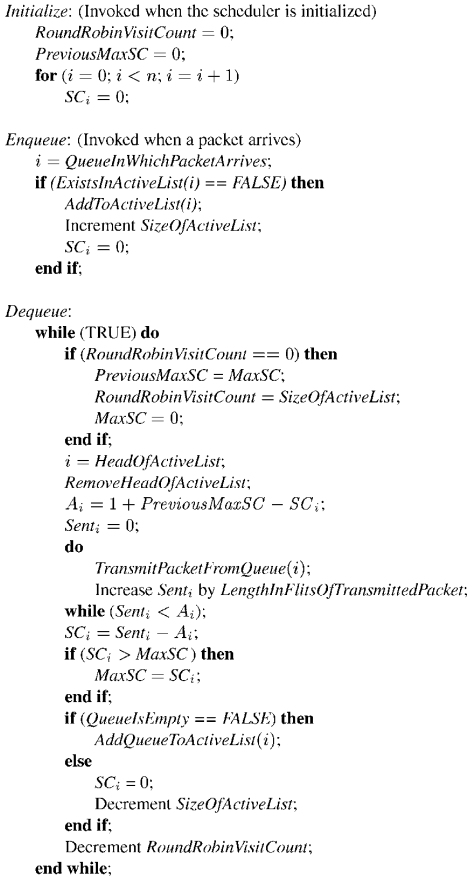}
\caption{Pseudo-code for ERR}
\end{figure}

Consider n flows, each with an associated queue with packets in it. The scheduler dequeues packets from these queues according to a scheduling discipline and forwards them for transmission over an output link. As in traditional scheduling problems, They allow that the length of time it takes to dequeue a packet is proportional to the size of the packet-however, to apply this work to wormhole networks, it is required that the scheduling algorithm not make any assumptions about the length of a packet prior to completely transmitting the packet. The ERR scheduler meets the above requirement. A pseudo-code implementation of the ERR scheduling algorithm is shown in Fig. 2, consisting of Initialize, Enqueue, and Dequeue routines. The Enqueue routine is called whenever a new packet arrives at a flow. The Dequeue routine is the heart of the algorithm which schedules packets from the queues corresponding to different flows. In their paper, they use a flit as the smallest piece of a packet that can be independently scheduled and they measure the length of a packet in flits.

They maintain a linked list, called the ActiveList, of flows which are active. A flow whose queue was previously empty and therefore not in the ActiveList is added to the tail of the list whenever a new packet belonging to the flow arrives. The ERR scheduler serves the flow i at the head of this list. After serving flow i, if the queue of flow i becomes empty, it is removed from the list. On the other hand, if the queue of flow i is not empty after it has received its round-robin service opportunity, flow i is added back to the tail end of the list.

Consider the instant of time, $t_1$, when the scheduler is first initialized. They define Round 1 as one round-robin iteration starting at time  $t_1$ and consisting of visits to all the flows that were in the ActiveList at time $t_1$. They illustrate this definition of a round using Fig. 3a. Assume that flows A, B, and C are the only flows active at the beginning of Round 1. The visits of the scheduler to flows A, B, and C, comprise Round 1. Let flow D become active after the time instant $t_1$, but before the completion of Round 1. Let the time instant $t_2$ mark the completion of Round 1. The scheduler does not visit flow D in Round 1 since D was not in the ActiveList at the start of Round 1. Round 2 is now defined as consisting of the visits to all of the flows that are in the ActiveList at time $t_2$. Assuming that flows A, B, and C are still active at time $t_2$, Round 2 will consist of visits to the flows A, B, C, and D.
\begin{figure}[ht]
\centering
\includegraphics[width=8cm]{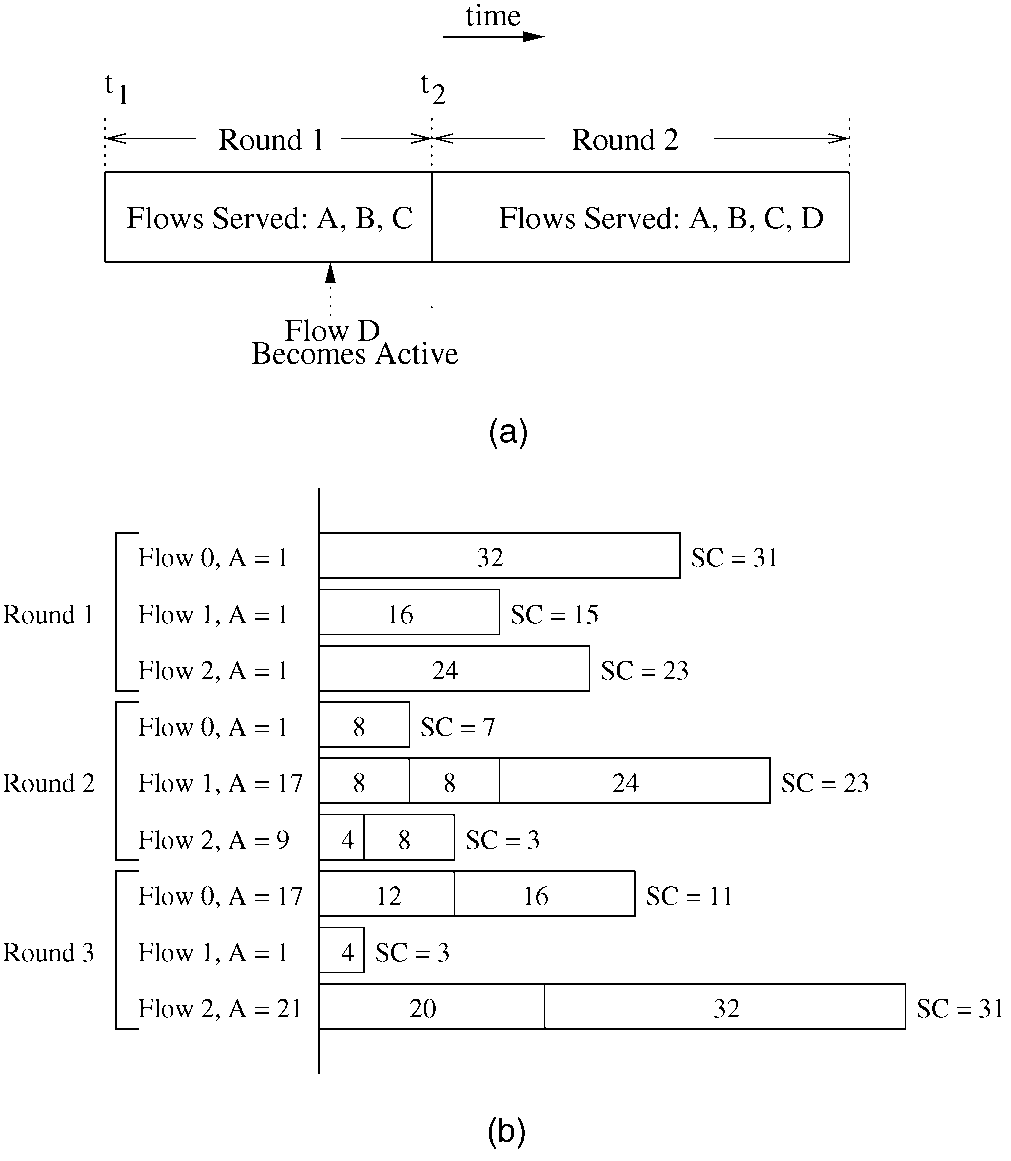}
\caption{(a) Definition of a Round and (b) an illustration of three rounds in an ERR execution}
\end{figure}

In general, they define round i recursively as the set of visits to all the flows in the ActiveList at the instant round (i - l) is completed. In order that the scheduler knows the number of flows it has to visit in any given round, they introduce the quantity RoundRobinVisitCount which denotes the number of flows that are in the ActiveList at the start of a round. RoundRobinVisitCount is decremented by one after each flow is served and, when it eventually equals zero, it implies the end of a round.

In each round，the scheduling algorithm determines the number of flits that a flow is allowed to send. They call this quantity the allowance for the flow during that round. The allowance assigned to flow i during round r is denoted by $A_i(r)$. This allowance, however, is not a rigid one and is actually elastic in that a flow may be allowed to send more flits in a round than its allowance. Let $sent_i(r)$ be the number of flits that are transmitted from the queue of flow i in round r. The ERR scheduler will begin serving the next packet from the queue if the total number of flits transmitted by the flow so far in the current round is less than its allowance. The ERR scheduler, thus, makes the scheduling decision without any knowledge about the packet length. Note that the last packet transmitted by a flow may cause it to exceed its allowance, as can happen when the allowance is smaller than the size of the packet at the head of the corresponding queue. When a flow ends up sending more than its allowance, it is interpreted as having obtained more than its fair share of the bandwidth. The scheduler records this unfairness in the Surplus Count (SC) associated with each flow. The surplus count, during any round, is the number of flits the flow sent in addition to its allowance. Let $SC_i(r)$ denote the surplus count of flow i in round r. Then, after serving flow i in round r, the scheduler computes $SC_i(r)$ as follows:
\begin{equation}
SC_i (r)= Sent_i(r)- A_i(r)
\end{equation}
Let MaxSC(r) denote the largest surplus count among all the flows served during round r. This quantity is used to recursively compute the allowances for each of the flows in the next round, using the following equation:
\begin{equation}
A_i(r)= 1+MaxSC(r-1)-SC_i(r-1)
\end{equation}
Note that, for the flow with the largest surplus count in the previous round, the new allowance is 1. This is ensured by the addition of 1 in (2) so that the scheduler will transmit at least one packet from this flow during the next round. The allowance given to each of the flows in a given round is not fixed and is computed depending on the behavior of the flows in the previous round. After the ERR scheduler serves flow i, if the queue of flow i is empty, its surplus count is reset to zero and it is removed from the ActiveList. Otherwise, if flow i has packets in its queue that are ready for transmission, it is added back at the tail end of the list.

Fig. 3b illustrates the first three rounds in an execution of the ERR scheduling discipline. In this figure, at the beginning of the first of these rounds, the surplus counts for all three flows and the MaxSC are all initialized to 0. Thus, from (2), the allowance during round 1 is equal to 1 for all the flows. The sizes of the packets actually sent by the flow during this round are shown by the horizontal bars and the new allowances for the next round are again computed using (1) and (2). It is easily observed from the figure that, in general, flows which receive very little service in a round are given an opportunity to receive proportionately more service in the next round.

\subsection{Eligibility-Based Round Robin}
The unfairness of ERR is due to the fact that surplus counters are reset as soon as flows become idle. This implies “forgetting” about any excess service received by idling flows, which are allowed to compete for the channel bandwidth again as early as in the subsequent round. The Eligibility-Based Round Robin (EBRR) scheduling algorithm is introduced by \cite{EBRR}. This approach avoids the “forgetting” flaw of ERR.
\begin{figure}[ht]
\centering
\includegraphics[width=6cm]{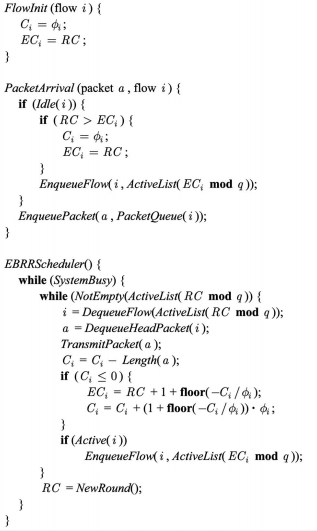}
\caption{Pseudo-code for EBRR}
\end{figure}

A pseudo-code implementation of the EBRR scheduling algorithm is shown in Fig. 4, consisting of FlowInit, PacketArrival and EBRRScheduler routines.

First of all, whenever an eligible flow is selected for service, instead of transmitting multiple packets back-to-back from that flow until either it is no longer eligible or the flow queue is emptied out, the flow is allowed to transmit one packet only. After that, if the flow is still eligible, it is inserted back into the current list so that every other eligible flow (if any) is allowed to transmit a packet before the flow can transmit again.

whenever a flow i becomes active after a packet arrival at an empty queue, the list into which the flow is inserted depends on how much surplus service, if any, the flow received the last time that it was active and completed the transmission of packets in its queue. The FlowInit function is invoked whenever a new flow is activated. It sets its credit to the quantum and its eligibility counter to the number of the current round, that is, the activating flow will be eligible as soon as it becomes backlogged. The PacketArrival function is invoked whenever a new packet arrives at the scheduler. If the flow was previously idle, it is now backlogged and has to be inserted into one of the active lists.

EBRRScheduler is invoked whenever there are packets to be scheduled. As long as the current list is not empty, the head-of-line flow i is selected for service. The head packet of flow i is dequeued and transmitted, and its length (after the packet transmission) is subtracted from flow i’s credit. When the current list is empty, a new round is started.

\subsection{Probabilistic Distance-based Arbitration}
Although round-robin arbitration provides local fairness across all routers, it does not provide any global fairness across all routers. As traffic flows through the network, it merges with newly injected packets and traffic from other directions in the network. This merging of traffic from different sources causes packets that have further to travel (more hops) to receive geometrically less bandwidth. As shown in Fig. 5, consider the 8-ary 1-mesh in Figure 5 where processors P0 thru P6 are sending to P7. The switch allocates the output port by granting packets fairly among the input ports. With a round-robin arbitration policy, the processor closest to the destination (P6 is only one hop away) will get the most bandwidth --- 1/2 of the available bandwidth. The processor two hops away, P5, will get half of the bandwidth into router R6, for a total of $1/2\times 1/2 = 1/4$ of the available bandwidth. That is, every two arbitration cycles P7 will deliver a packet from source P6, and every four arbitration cycles it will deliver a packet from source P5. As a result, P0 and P1 each receive only 1/64 of the available bandwidth into P7, a factor of 32 times less than that of P6.
\begin{figure}[ht]
\centering
\includegraphics[width=11cm]{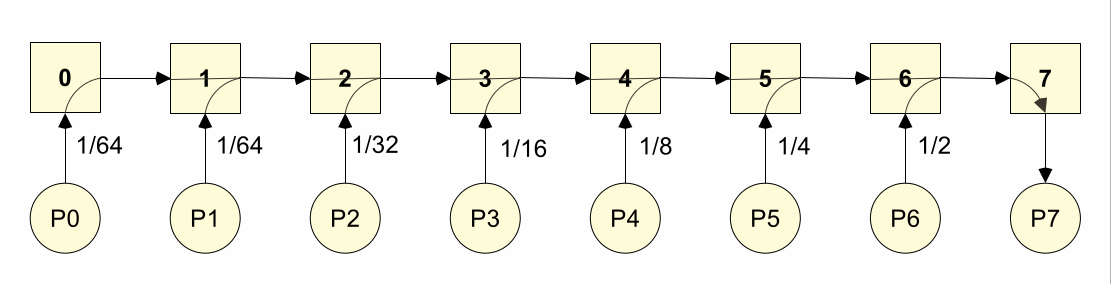}
\caption{8-ary 1-mesh example where all nodes are sending to P7 and merging traffic at each hop}
\end{figure}
Age-based arbitration \cite{Age-based} is known to provide global fairness as when two or more packets arbitrate for a shared resource, the packet with the oldest age wins the arbitration. However, it become complex to implement in an on-chip network constraint.
EoS \cite{EoS} is proposed to provide global fairness without any significant complexity and relying only on local arbitration. 
EoS uses probabilistic distance based arbitration, thus avoiding the complexity with age-based arbitration by proposing to approximate the age of a packet with distance or hop count. By using information already present in the packet, such as source node, current node, or destination node and using distance as a proxy for the packet’s age, age-based arbitration is greatly simplified.

By using priority based on hop count, livelock and fairness issues are problematic because packets with a lower priority (i.e., a lower hop count) can continually lose arbitration because of a constant stream of newly injected traffic with higher priority. To overcome this problem while still using hop count as the weight, EoS propose probabilistic arbitration where the output of the arbitration is probabilistically determined based on the weight of the input requests.
Assume an arbiter with m requests, for each requests ri with a corresponding weight $w_i$. The probability of $r_i$ being granted is 
\begin{equation}
P(g_i) = \frac{w_i}{\sum_{j=1}^{j=m}w_j}
\end{equation}
As shown earlier in Figure 5, nodes that are farther away are serviced at a rate that is exponentially proportional to the hop count. To account for this difference, EoS introduce nonlinear weights based on the distance. Instead of using a weight which is equal to the hop count (i.e., w = h), they introduce nonlinear weights in probabilistic arbitration , i.e. $w = C^h$ where C is the contention degree or the number of packets contending for the same output port. By using nonlinear weights, better fairness is provided for nodes that are farther away.

Three arbitration metrics are designed in EoS, with C and h to be static or dynamic.
\begin{enumerate}[1)]
\item Fix weight. The total number of hops a packet must travel from its source to its destination is a static value in a mesh network with minimal routing. The exponential h is set to be distance from source to destination, and C is set to be max contention degree in each router.
\item Constantly Increasing Weight. the exponential h is set to be the distance a packet has traversed, and C is set to be max contention degree in each router.
\item Variably Increasing Weight (VW). Instead of assuming a constant C value at each hop in each dimension, C is set dynamically to be the actual contention degree.
\end{enumerate}
On hotspot traffic where all nodes send traffic to a single destination, EoS match the performance of age-based arbitration in hotspot traffic and achieve equality of service.

\section{Motivation}
\subsection{The drawback of RFB}
In wormhole switch networks, the contention resources refer to the buffer, channel and crossbar. All the algorithms mentioned above measure the fairness based on the allocation of crossbar as the total length of packets forwarded in each flow is nearly identical. While contention and congestion always exist in the network, knowing the size of a packet does not imply we have the information that how long it will hold the channel, a.k.a. service time, since the latter might include blocking time.

RFB merely consider the metric of crossbar fairness, which may result in the decrease in the throughput of the network. We will conceive an instance to illuminate this side-effect. Assume we have two flows A and B, and A is congested. As the granularity of scheduling is packet, when A is granted, it will cost a host of cycles to forward this whole packet and worsen the congestion at the same time. Then B is to some extent a victim as its latency soars due to A’s blocking time. By definition, in the next round the grant will allocate to A and lead to a vicious circle. However, this situation could be mitigated significantly if we take the congestion of flow A into account. When we perceive the service time is much longer than its packet size, we assert that this flow face the congestion. We will reduce the priority of this flow and give more chances for the other free-congestion flows. Evidently the schedule violates RFB and is classified into unfair algorithm, while it is really reasonable and has the ability to achieve high throughput and low latency.
\subsection{Channel Fairness Bound}
The drawback of RFB is that it only takes crossbar resource as the metric and ignores the occupation time of the channel. We propose a novel evaluation Channel Fairness Bound (CFB) to evaluate the fairness of packet scheduling algorithm in a bid to evade the flaw of RFB. RFB merely comprises sending time, i.e., the packet size, while CFB consists of sending time and blocking time. 
There are two methods to make use of blocking time. We can still try to follow the evaluation standard of RFB to keep the sending time fair. However, we should set a threshold to detect the congestion and further modify the priority or the round-robin pointer. When the service time of a certain flow is over the threshold, it means congestion occurs and it is unadvisable to allocate the resources as usual. Free-congestion flows tend to receive the grants. But this modification is temporary. We presume this traffic jam will relieve significantly in the next several round and the flow can restore its previous status. It is worth noting that during the time the flow is degraded, the allocated crossbar resources go down accordingly. Whether a compensation is necessary depends the concrete algorithm.

Another way is to measure fairness with sending time and blocking time instead of simple sending time. As briefly introduced above, DRR, ERR and EBRR can easily fulfil the requirement of CFB. We just replace the packet size with channel occupation time and the fairness is the same as in the RFB. While as J. Kim et al. \cite{EoS} point out, round-robin arbitration provides local fairness across all routers, it does not provide any global fairness across all routers. It is likewise a challenge if we adopt CFB. Here we will use the similar method with probabilistic distance based arbitration to achieve global fairness.
\subsection{Probabilistic Arbitration}
For convenience, we derive the theoretical feasibility of CFB to achieve global fairness on hotspot traffic where all nodes send traffic to a single destination. Probabilistic distance based arbitration guarantees equality of service by measuring the accepted throughput across all nodes. The metric of fairness is still RFB. Similarly, we will test whether this algorithm is capable to provide global fairness on the basis of CFB.

A hotspot traffic example is shown in Fig. 6 where processors P0 through P6 are sending data to P7. Generally speaking, the packet size is fixed in a specific network. So let L be the length of all packets.
\begin{figure}[!ht]
\centering
\includegraphics[width=11cm]{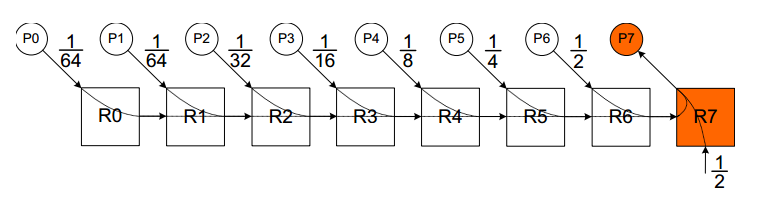}
\caption{8-ary 1-mesh example where all nodes are sending to P7}
\end{figure}

We define some notations here. Let $T_{ij}$ be the VC service time which router j offers to flow i , $t_{ijm}$ be the VC service time that router j offers to packet m of flow i , $k_{ij}$ be the number of packets in flow i going through router j. By definition, we have
$T_{ij} = \sum^{k_{ij}}_{m = 1}t(ijm)$
We denote by $t_{ij}$ the average VC service time a packet in flow i is allocated. Then there is 
\begin{equation}
T_{ij} = k_{ij} t_{ij} 
\end{equation}

From equation (4), we know that to ensure equality on VC service time for each flow, $k_{ij}$ should be in inverse proportion to $t_{ij}$, i.e., the number of packets in flow i  should be in inverse proportion to VC service time router j offers to a packet in flow i. As in probabilistic distance based arbitration, the weight of each flow determines $k_{ij}$, therefore the only problem to satisfy CFB is to attain the information of $t_{ij}$ and further adjust $k_{ij}$ accordingly. This method can ensure that the channel occupation time in the destination node for each flow is nearly the same, which is one of the metric of CFB. While probabilistic distance based arbitration also achieves the global fairness under RFB in each node besides the destination, we will explore whether this scheduling can guarantee fairness under CFB.
Take $W(ij)$ as the weight of flow i in router j , $R_j(i)$ as the number of packets j accepts from flow i , and n as the number of flows router j accepts. By definition, there is 
\begin{equation}
R_j = R_j(0):R_j(1):R_j(2):\dots:R_j(n)
\end{equation}

Trivially, we can obtain
\begin{equation}
R_1 = W(01):W(11)
\end{equation}

Consider R0 and R1 in Figure 6. Flow 0 consists of packets whose source is P0. When flow 0 arrives at R1, its weight is $W(01)$. When flow 1 arrives at R1, its weight is $W(11)$. According to the relation of weight and the number of packets described above, there is equation (6).
By the same token, we can get
\begin{equation}
R_2=W(01):W(11):(\frac{W(22)}{W(02)}W(01)+\frac{W(22)}{W(12)}W(11))
\end{equation}

Consider R0, R1 and R2 in Figure 6. When flow 0 and 1 arrives at R1, the weight proportion is (W(01), W(11)) . Then flow 2 arrives at router 2. To maintain the weight proportion of flow 0, 1, 2 at router 2, when flow 0 sent W(01) packets and flow 1 sent W(11) packets, number of packets flow 2 can send should be $\frac{W(22)}{W(02)}W(01)+\frac{W(22)}{W(12)}W(11)$. Then there is equation (7)
In general, we have
\begin{equation}
\begin{split}
R_1=W(01):W(11) \\
R_j=R_{j-1}: \sum^{j-1}_{i=0} \frac{W(jj)}{W(ij)}R_{j-1}(i) \\
\end{split}
\end{equation}

For simplicity, let W(jj) = 1. So equation is then simplified as follows
\begin{equation}
R_j = R_{j-1}: \sum^{j-1}_{i=0}\frac{1}{W(ij)}R_j-1(i)
\end{equation}

We observe that packets that go through router j-1 will also go through router j.
Assume that the service time for all flows in each router is under equality of service. We denote by $S_{ij}$ the ratio of channel occupation time to sending time. With $R_1=W(01):W(11)$, we have
\begin{equation}
W(01)S(01)=W(11)S(11)
\end{equation} 

In a similar way, we can get
\begin{equation}
\frac{w(01)}{w(11)} = \frac{s(12)}{s(02)}
\end{equation}

Therefore, we can obtain the equality without W
\begin{equation}
\frac{s(11)}{s(01)} = \frac{s(12)}{s(02)}
\end{equation}

In general, we have 
\begin{equation}
\frac{S(mk)}{S(nk)} = \frac{S(mt)}{S(nt)}
\end{equation}

In wormhole switching networks, restriction (13) cannot be satisfied. Thereby probabilistic distance based arbitration cannot achieve fairness under CFB.

\bibliography{reference}
\end{document}